\documentclass[12pt]{iopart}

\usepackage{iopams}  
\usepackage{graphicx}
\usepackage{dcolumn}
\usepackage{bm}
\usepackage{xfrac}
\usepackage{braket}
\usepackage{wrapfig}
\usepackage{subcaption}
\usepackage[makeroom]{cancel}
\usepackage{hyphenat}
\begin{document}

\title[Delayed choice experiment using atoms in optical cavity]{Delayed choice experiment using atoms in optical cavity}

\author{Sankaranarayanan Selvarajan}

\address{Tata Institute of Fundamental Research (TIFR), Mumbai, India}
\ead{sankar.s@tifr.res.in}
\vspace{10pt}
\begin{indented}
\item[]\today
\end{indented}

\begin{abstract}
In this article, we propose a method to realize the ``delayed choice experiment" using ultra-cold atoms. Here we attempt to probe the ``welcher-Weg" information without collapsing the wavefunction of the atom. This experiment consists of components built around proven techniques that are put together in novel configuration to preserve the coherence of the system during the measurement. The Ramsey interference is used to establish the wave nature of the atom and the particle nature of the atom is probed by detecting its internal state by performing a non-demolition measurement using an ultra-high finesse cavity. The coherence of the atom is preserved by adjusting the atom-cavity interaction time such that the state of the atom is unchanged when it emerges out of the cavity.
\end{abstract}

\maketitle

\section{Introduction}

The principle of complimentary formulated by Niels Bohr \cite{bohr1928quantum}, which forbids the simultaneous observation of particle and wave nature is perhaps the most fundamental aspect of quantum physics \cite{feynman1965feynman}, which distinguishes it from the intuitive realm of classical mechanics. This complimentary nature is manifested as the quantum superposition i.e., state of a physical quantum system can be expressed as a superposition of other states. Quantum delayed choice experiment \cite{Wheeler1983quantum} emerged as one of the best methods to probe this intriguing aspect of QM in the laboratory.

In the simplest picture, for this gedanken experiment, the quantum particle is sent through a Mach-Zender interferometer. The length of the interferometer arms is adjusted such that the particle always emerges from the same output port, due to constructive interference. This underscores the wave nature of the quantum particle. However, any attempt made to find out which path the particle took while it is traveling through the interferometer, destroys the interference and the probability of finding the particle in both the output ports becomes equal, i.e., it behaves like a particle. This experiment is called ``Welcher-Weg" (which way) measurement. The choice of performing the which-way measurement can be made after the particle is already inside the interferometer, thus ruling out the possibility of particle sensing in advance, the kind of experiment it is going to encounter. This is known as a delayed choice experiment,

While many groups have demonstrated the delayed choice experiment over the years \cite{hellmuth1987delayed, durr1998origin, kim2000delayed, jacques2007experimental, manning2015}, it has been a general concern that the coherence of the quantum system is fragile and any interaction leading to the which-path measurement destroys the interference. For instance, a common method to perform the which-path measurement is by mapping the path information onto the internal state of the system. This mapping projects the state of the system onto its orthogonal basis, which invariably leads to the loss of interference. This introduces an ambiguity in explaining if the interference is lost because we now possess the which-path information, or because the state of the systems has been altered in the process of acquiring the which-path information.

In this proposal, we attempt to overcome this problem by preserving the coherence of the atom using non-demolition measurement. The wave nature of the atom is demonstrated by the Ramsey internal state interference between two the long-lived Rydberg states of the atom and an ultra high finesse cavity coupled to a different transition of the atom is used to probe the particle nature of the atom by detecting its internal state. During the measurement, the state of the atom is preserved by adjusting the atom-cavity interaction time such that the atom completes a 2$\pi$ rotation due to vacuum Rabi oscillations. Strictly speaking, this would be an ``which-state" experiment, however, the internal state of the atom can be entangled to its path by using state-dependent light-atom interaction \cite{durr1998origin} and convert this into a true ``which-path" experiment. This proposal relies on independent techniques used for atom interference and state detection that has been successfully demonstrated by various groups. Delayed choice experiments using QND measurements have been proposed before \cite{storey1993atomic, gerry1996complementarity}. However, they are based on entirely different configurations.

\section{Scheme}
\label{sec:scheme}

 \begin{figure}[htp]
    \centering
    \includegraphics[width=8.6cm]{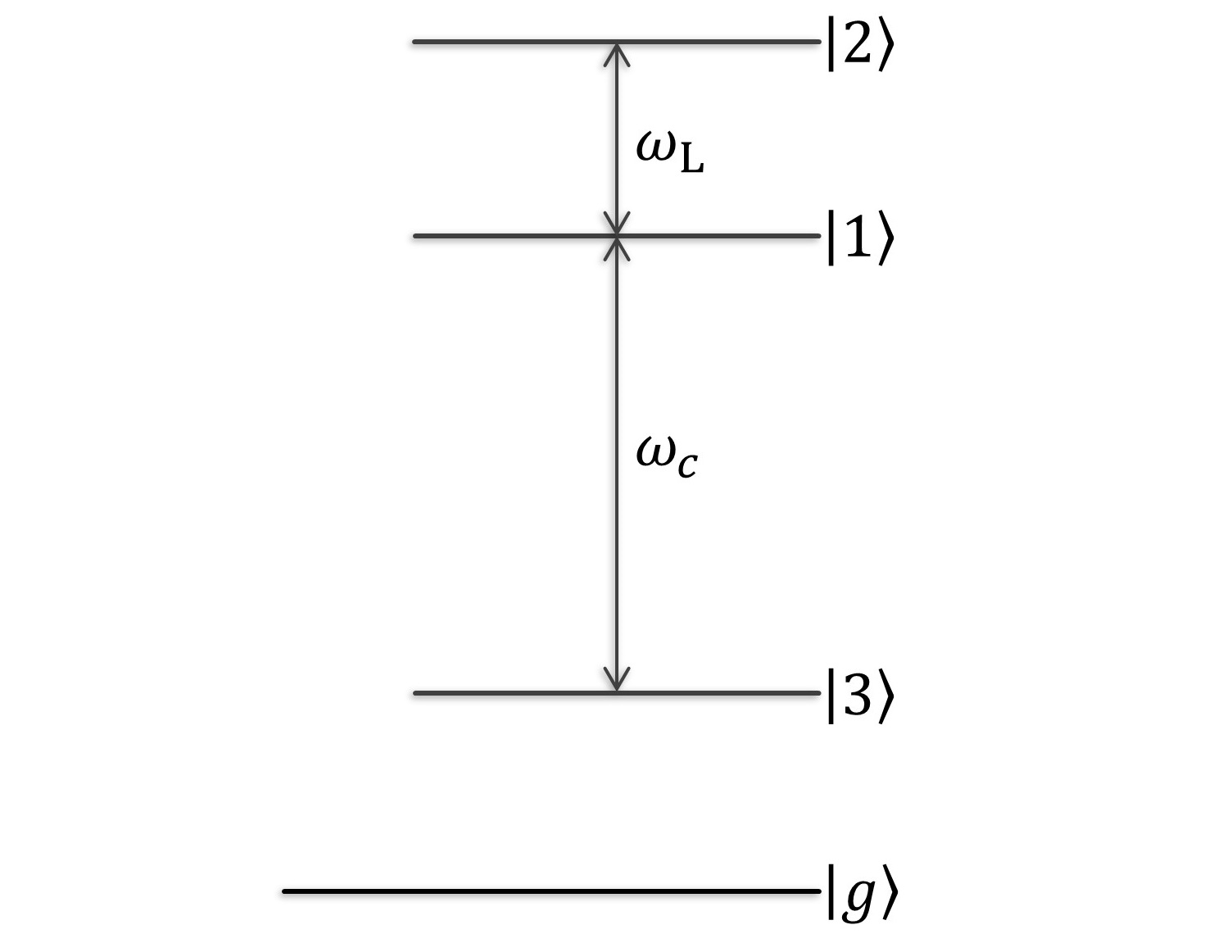}
    \caption{Energy level of the atom required to carry out the experiment. a Ramsey interference using states $\ket{1}\&\ket{2}$ establishes the wave nature of the atom. ``Which way" information is probed using the non-demolition measurement on $\ket{1}\rightarrow\ket{3}$ transition.}    
    \label{fig:atom_level}
\end{figure}

\begin{figure*}[htp]
    \centering
    \includegraphics[width=.8\textwidth]{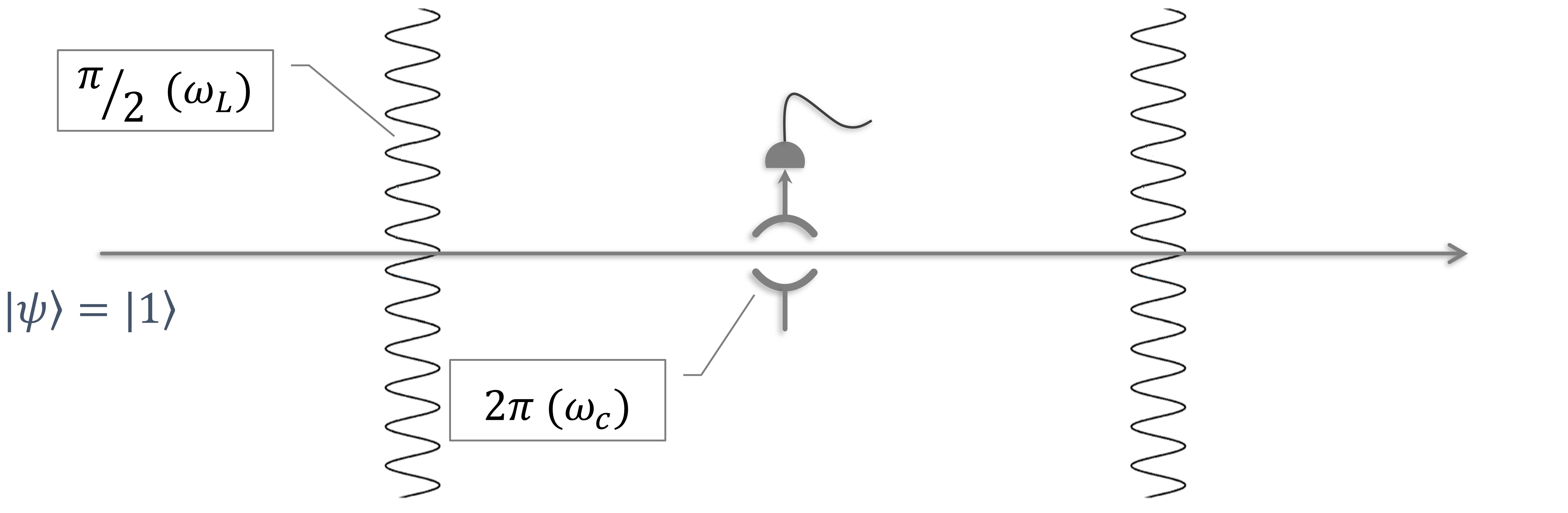}
    \caption{Schematic of the experiment: Here the atom is travelling from left to right with initial state $\ket{1}$. The two $\pi/2$ pulses enable the Ramsey interference between $\ket{1}\&\ket{2}$. An ultra high finesse cavity after the first $\pi/2$ pulse, tuned to $\ket{1}\rightarrow\ket{3}$ transition probes if the atom is in state $\ket{1}$ or not, without destroying its coherence.}
    \label{fig:expt_scheme}
\end{figure*}

The atomic-level structure suitable for carrying out the experiment is shown in Fig.~\ref{fig:atom_level}. The wave nature of the atom is established using the Ramsey internal state interference as demonstrated in the reference \cite{gleyzes2007quantum}. consider the atoms are initially in state $\ket{1}$. A $\pi/2$ pulse resonant with $\ket{1}\rightarrow\ket{2}$ transition puts the atom in a superposition of state $\ket{1}$ and $\ket{2}$. The second $\pi/2$ pulse enables the internal state interference resulting in a final state of the atom depending on the relative phase acquired by the atom between the two $\pi/2$ pulses. The separation between the $\pi/2$ pulses is adjusted such that the final state of the atom after interference is $\ket{1}$.

In our proposal, we probe the particle nature of the atom by detecting its internal state. This involves a method similar to the one used in reference \cite{Hood1998real}, to detect atoms in real-time using an ultra-high finesse cavity in the Fabry–Perot configuration. The frequency of this cavity is tuned to be in resonance with the $\ket{1}\rightarrow\ket{3}$ transition of the atom ($\omega_c = \omega_{1\rightarrow 3}$). A weak prob laser beam resonant with the empty cavity ($\omega_L = \omega_c = \omega_{1\rightarrow 3}$) is aligned to the optical input port of the cavity. This probe beam can be switched on and off with high temporal precision. The beam transmitted through the cavity is detected using a photodetector.

Consider an atom inside the cavity. If it is in the state $\ket{2}$, the cavity and the detection laser are detuned from the atomic transition by $\omega_{1\rightarrow 2}$. This suppresses any strong interaction between the atom and the probe laser, and no change is observed in the photodetector signal at the cavity output port. We denote this with $\ket{\uparrow}$ representing the high state of the photodetector. This interaction results in a shift in the phase of the atom, which will be addressed in the subsequent sections. If the atom is in state $\ket{1}$ then it gets coupled to the cavity and the resonance frequency gets split into two in a phenomenon known as vacuum Rabi splitting. This makes the detection laser off-resonant to the cavity and the atom, which results in its destructive interference inside the cavity. This ensures that no photons of prob beam enter the cavity and consequently the photodetector reads a low signal on the output port of the cavity. we denote this with $\ket{\downarrow}$. Using this method we should in principle be able to detect if the atom is in state $\ket{1}$ or $\ket{2}$ while preserving its coherence.

\paragraph*{``Welcher-Weg" experiment:}
In our proposal, the ``welcher-Weg" experiment is accomplished by combining the above interference and detection mechanisms. The atom is prepared in a superposition of $\ket{1}$ and $\ket{2}$ states and made to pass through the high finesse cavity tuned to $\ket{1}\rightarrow\ket{3}$ transition. The prob laser is turned on when the atom is inside the cavity. This operation projects the state of the atom onto the state of the photodetector non-destructively. Finally, the Ramsey interference is completed by shining in the second $\pi/2$ pulse and the final state of the atom is measured. The aim of this experiment is to show that any attempt made to know the state of the atom would destroy the interference of the final state. This is accomplished by tagging the final state of the atom to the corresponding photodetector state measurement. This in principle should result in the destruction of the interference of the atomic state. A detailed analysis of the state of the system at various points in the experiment is presented below. 

The schematic of the experiment is shown in Fig.\ref{fig:expt_scheme}. Independent atoms optically pumped into state $\ket{1}$  are launched into the interference setup. The first $\pi/2$ pulse places the atoms in a superposition state and the wave function can be written as

\begin{equation} \label{eqn_1}
\ket{\psi} = \frac{1}{\sqrt{2}}\big(\ket{1}+i\ket{2}\big).
\end{equation}

The atoms are then made to transit through the high finesses cavity. The atom-cavity interaction time has to be adjusted such that the atom undergoes N$\times2\pi$ Rabi oscillations due to the interaction \cite{assemat2019quantum}. This ensures that the state of the atom is preserved. The state of the atom is detected non-destructively in real-time using the probe laser beam. The probe beam is switched on when the wave function of the atom is maximally coupled to the cavity, i.e., when the atom is close to the center of the cavity. The duration of the probe laser is adjusted to be much smaller than the transit time of the atom through the cavity, but long enough to keep the linewidth of the laser smaller than the Vacuum Rabi splitting. This process entangles the internal state of the atom with the state of the detector and results in a so-called ``Schr\"{o}dinger's cat" state or ``Quantum switch" \cite{davidovich1993quantum}. The state of the system including the atom and the detector is written as

\begin{equation} \label{eqn_2}
\ket{\psi} = \frac{1}{\sqrt{2}}\big(\ket{1}\ket{\downarrow}+ie^{i\Theta}\ket{2}\ket{\uparrow}\big),
\end{equation}

where $\Theta$ is the total phase difference acquired by two states of the atom between the two $\pi/2$ pulses. This comprises of $\theta$, the usual phase difference of the Ramsey interference, and $\phi$, phase difference acquired due to the introduction of the probe cavity. This includes the phase introduced by a) the off-resonance laser light on state $\ket{2}$ and b) the Vacuum Rabi oscillations driven by the cavity on the $\ket{1}\rightarrow\ket{3}$ transition.

After the second $\pi/2$ pulse the wavefunction of the system can be written as 

\begin{equation} \label{eqn:DC_config}
\ket{\psi{}} = \frac{1}{2}\bigg(\Big(\ket{1} + i\ket{2}\Big)\ket{\downarrow} + ie^{i\Theta}\Big(i\ket{1} + \ket{2}\Big)\ket{\uparrow}\bigg).
\end{equation}

If we consider only the atoms that trigger the cavity, i.e. when the state of the photodetector is $\ket{\downarrow}$. Then the wavefunction of the system is written as 

\begin{equation}
 \ket{\psi_\downarrow} = \frac{1}{2}\Big(\ket{1} + i\ket{2}\Big)\ket{\downarrow},
\end{equation}

where $\ket{\psi_i} = \ket{i}\braket{i|\psi}$. Here, the probability of finding the atom in the state $\ket{1}$ or $\ket{2}$ is independent of the phase difference acquired and the interference is lost.

\begin{equation}
 |\psi_1|^2 = |\psi_2|^2  = \frac{1}{2} |\psi_\downarrow|^2.
\end{equation}

This loss of interference is due to the correlation of the state of the atom to the photodetector signal and need not be due to the decoherence of the atom.

\begin{figure}
     \centering
     \begin{subfigure}[b]{8.6cm}
         \centering
         \includegraphics[width=\textwidth]{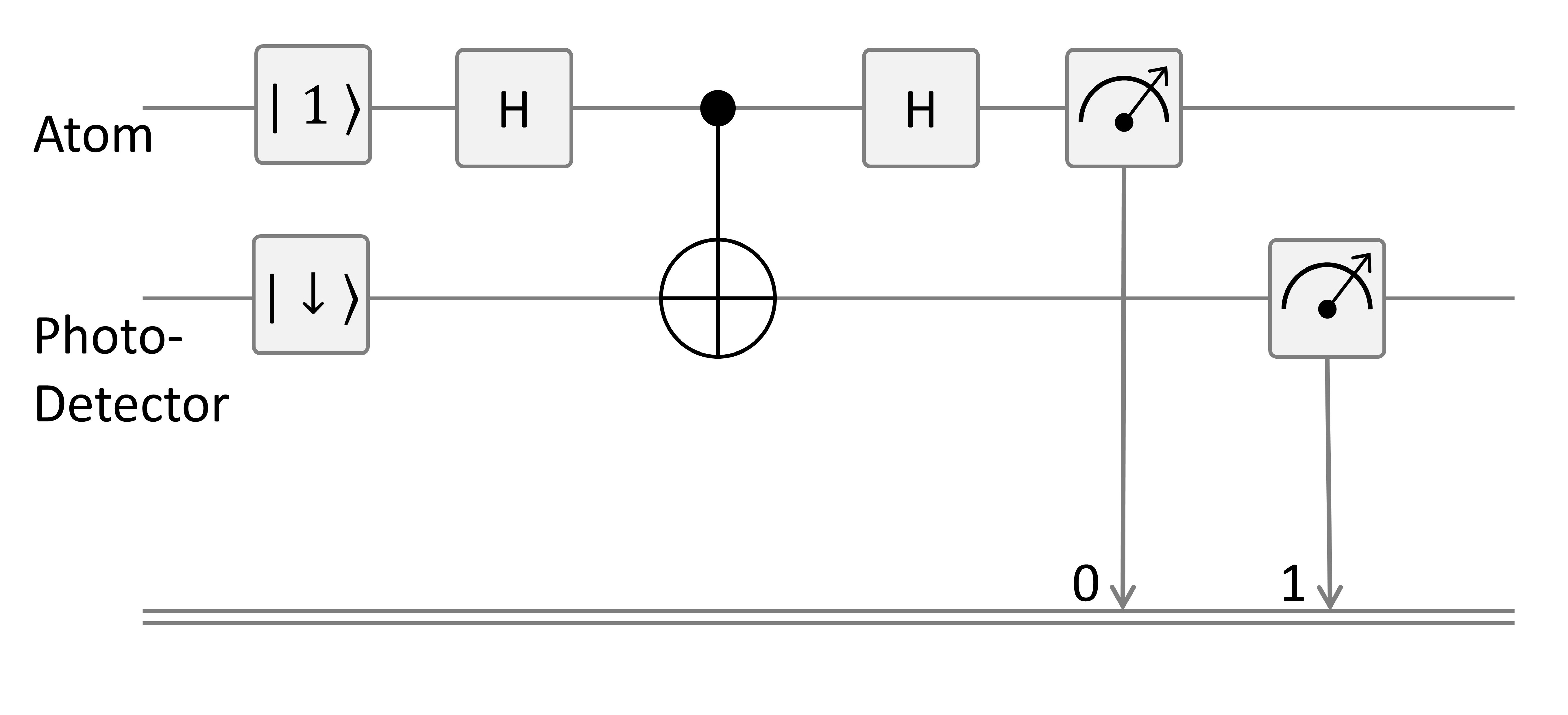}
         \caption{}
         \label{fig:Qcirc}
     \end{subfigure}
     \\
     \begin{subfigure}[b]{8.6cm}
         \centering
         \includegraphics[width=\textwidth]{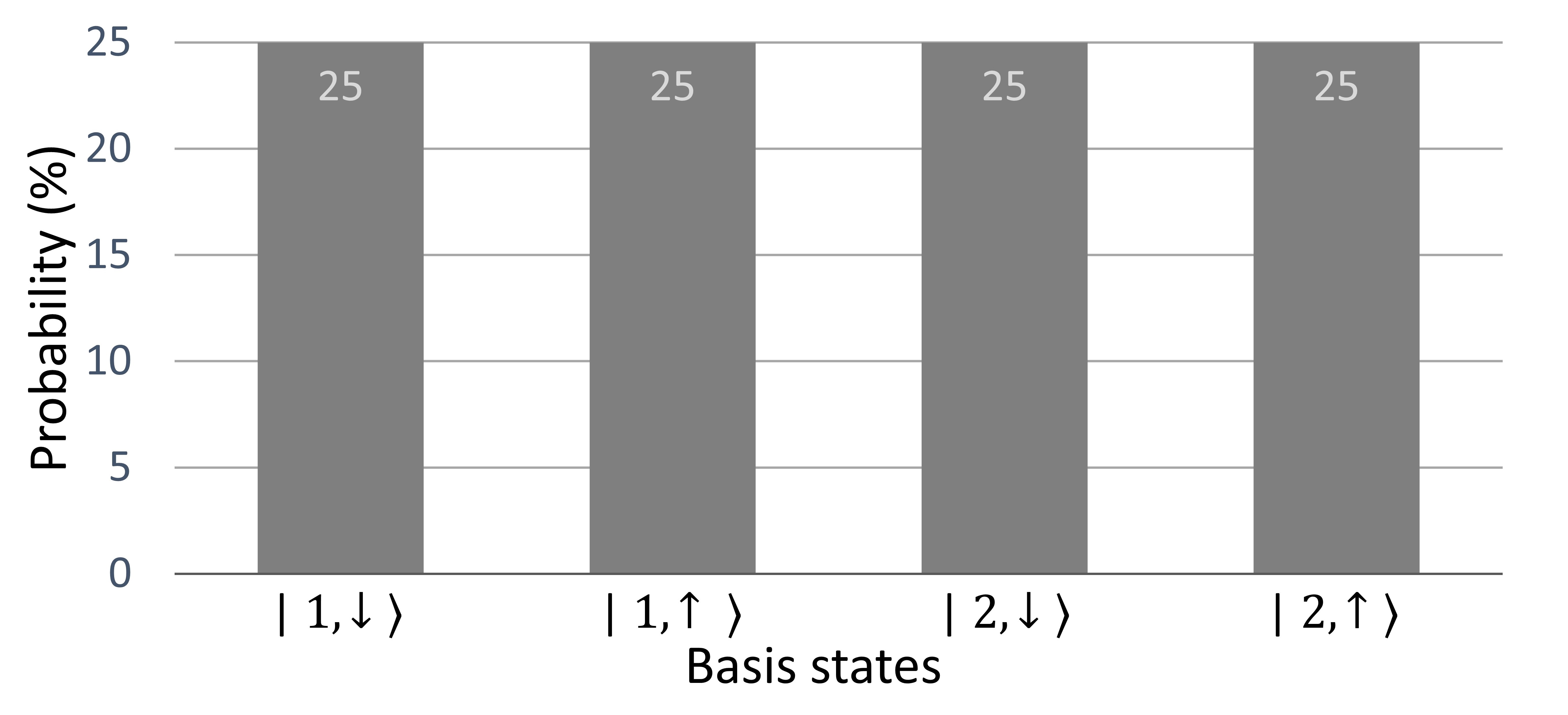}
         \caption{}
         \label{fig:probs}
     \end{subfigure}
        \caption{a)~Quantum circuit representation of the experiment. The atom is prepared in state $\ket{1}$ while the detector is initially at a lower state $\ket{\downarrow}$. The $\pi/2$ pulses are represented by the Hadamard gate H. The atom-cavity interaction is essentially a controlled-NOT gate, where the state of the cavity is dependent on the state of the atom. b)~Plot showing the probabilities of finding the system in each of the final states of the atom and the photodetector, suggesting the loss of interference.}
        \label{fig:Qcircuit}
\end{figure}

The quantum circuit type representation of the process is shown in Fig.~\ref{fig:Qcircuit}. Initially, the atom and the detector are created in the states $\ket{1}$ and $\ket{\downarrow}$ respectively. The first local Hadamard gate on the atom maps its state to an equal superposition state of $\ket{1}~\&~\ket{2}$. The atom-cavity interaction is essentially a controlled-NOT gate on the state of the detector based on the state of the atom. This operation entangles the state of the detector to the state of the atom. Finally, the second Hadamard gate on the atom concludes the sequence of Ramsey interference, followed by the measurements on the state of the atom and the photodetector. The probabilities of finding the atom-cavity system in each of the states is plotted in Fig.~\ref{fig:probs}. For a given state of the detector, the probability of finding the atom in the states $\ket{1}$ or $\ket{2}$ is the same, suggesting that the interference is lost.

In the situation where the ``welcher-Weg" information cannot be obtained, i.e., when the probe laser is not switched on, the final state of the system is written as

\begin{equation}\label{eqn:QE_config}
\ket{\psi{}} = \frac{1}{2}\bigg(\Big(\ket{1} + i\ket{2}\Big) + ie^{i\Theta}\Big(i\ket{1} + \ket{2}\Big)\bigg),
\end{equation}

and the interference survives. 

\begin{equation}\label{eqn:in_pro_withoutC}
 |\psi_1|^2 = \frac{1}{2} (1-\cos{\Theta}),\nonumber\\
 |\psi_2|^2 = \frac{1}{2} (1+\cos{\Theta}).
\end{equation}

Thus establishing the complementary wave-particle nature of the atoms.

\paragraph*{Delayed choice experiment:}
To operate this experiment in the ``Delayed choice" configuration, the choice of probing the particle nature of the atom should be made after the atom is placed in the superposition state. This can be achieved by switching on the atom-cavity interaction after the first $\pi/2$ pulse. A randomly generated binary signal dictates the choice of atom-cavity interaction with equal probability. After the second $\pi/2$ pulse the final outcome of the experiment can be analyzed with or without the atom-cavity interaction simply by co-relating the binary signal to the state of the atom.

\section{Experimental procedure}

 \begin{figure}
    \centering
    \includegraphics[width=8.6cm]{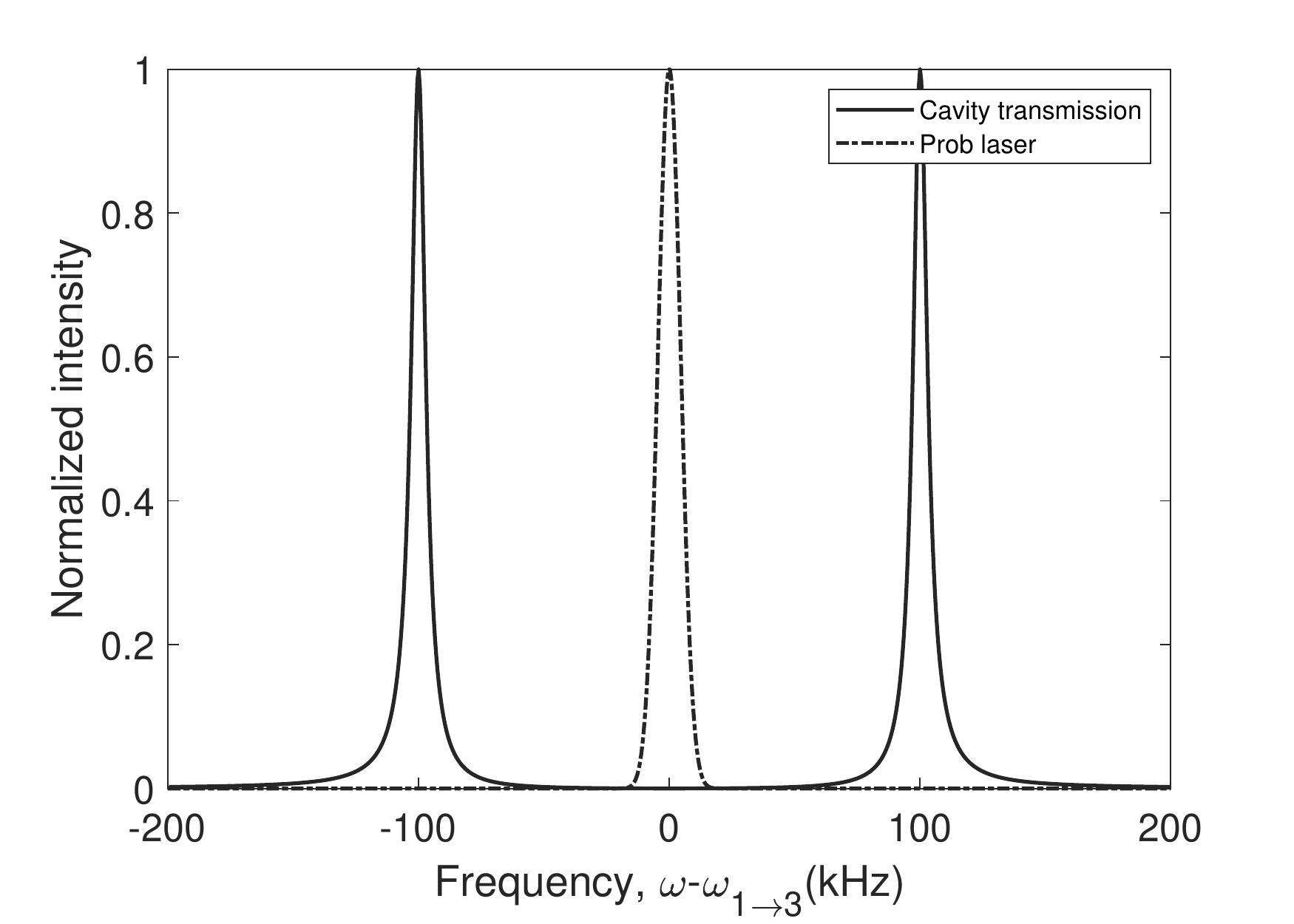}
    \caption{Theoretical model showing the vacuum Rabi splitting of the coupled atom-cavity system compared to the spectral width of the probe laser pulse. The solid curve shows the transmission spectrum of the maximally coupled atom-cavity system with typical values of  ($g,\kappa,\Gamma)/2\pi = (100,5,2)$ kHz. The dashed curve depicts the spectral profile of the 35~$\mu s$ probe laser pulse. Both curves are normalized and the origin of the X axis coincides with $\omega_{1 \rightarrow 3}$.}
    \label{fig:Cavity_trans}
\end{figure}

An experimental setup similar to the one used in the reference \cite{gleyzes2007quantum}, with some modifications is well suited to carry out this experiment. The states $\ket{1}$ and $\ket{2}$ mentioned in Fig.~\ref{fig:atom_level} can be realized using the Rubidium atom in long-lived circular Rydberg levels with principal quantum numbers n~=~50 and n~=~51 respectively. The atoms are velocity selected and optically pumped into state $\ket{1}$. Two low finesse microwave cavities, resonant with $\ket{n=50}\rightarrow\ket{n=51}$ transition is used for the $\pi/2$ pulses. The final state of the atom is detected using a state-sensitive field ionization detector. The distance between the two low finesse Ramsey cavities is adjusted such that the atom ends up in a $\ket{n = 50}$ state after the interference.

For probing the internal state of the atom between the two $\pi/2$ pulses, we need an optical cavity in resonance with an atomic transition connecting the state $\ket{n=50}$ of the atom with a third level (corresponding to $\ket{3}$ in Fig.~\ref{fig:atom_level}). This level is chosen such that its spontaneous decay rate is much smaller compared to the atom-cavity coupling while keeping the wavelength of the transition accessible with commercial lasers. n = 11, L = 10 could be one such level.  The typical lifetime of the atom in n = 11 orbit is of the order of 80$~\mu s$ \cite{sansonetti2006wavelengths}. By choosing the circular Rydberg level this can be further increased to suit the needs of the experiment. The $\ket{1}\rightarrow\ket{3}$ transition frequency, $\omega_{1\rightarrow 3}$ is close to 6.5~$\mu m$, within the reach of commercial lasers.


Real-time non-destructive detection of the atomic state is carried out using the method demonstrated in the reference \cite{Hood1998real}. The ultra-high finesse cavity consists of two spherical super-polished mirrors. The parameters of the atom-cavity system are adjusted to be ($g,\kappa,\Gamma)/2\pi= (100,5,2)$ kHz, where $g$ is the coherent atom-cavity coupling dictated by the geometry of the cavity, $\kappa$ linewidth of the TEM$_{00}$ mode of the cavity which depends on the reflectively of the mirrors, $\Gamma$ is the combined decay rate of the levels $\ket{1}$ and $\ket{3}$ of the atom, which in this case is limited by the lifetime of level $\ket{3}$.

The resonant frequency of the empty ultra-high finesse cavity ($\omega_c$) and the frequency of the probe laser ($\omega_L$) is stabilized to be resonant with the $\ket{1}\rightarrow\ket{3}$ atomic transition while keeping the linewidth of the probe beam $\Gamma_L \le \kappa$. The probe laser beam is aligned such that the TEM$_{00}$ mode of the cavity is excited. The transmission of the probe laser with typical powers of 100~pW is measured with a balance heterodyne detection as demonstrated in \cite{Hood1998real}. The pulse width of the probe laser T$_p$ should be around 35~$\mu$s resulting in a linewidth of about 30~kHz, sufficiently small compared to the vacuum Rabi splitting. The shape of the probe pulse can be modified to form a Gaussian for better suppression of frequency components away from the center frequency $\omega_L$. A comparison between the vacuum Rabi splitting and linewidth of the probe laser pulse is shown in Fig.~\ref{fig:Cavity_trans}. The probe beam is switched on when the atom is maximally coupled to the cavity, i.e., when the atom is close to the center of the cavity. This guarantees that the coupled atom-cavity energy levels are furthest apart and the atom is not excited by the probe laser pulse.

To minimize the time atom spends in level $\ket{3}$, the atom-cavity interaction time T$_i=\text{N}\times\text{T}_\Omega$ is chosen to be around twice the duration of the probe laser pulse, T$_p$. Here T$_\Omega$ is its time-period of vacuum Rabi oscillations. The atom-cavity interaction is controlled by stark shifting the energy level of the atom by applying an electric field across the cavity \cite{assemat2019quantum}. The cavity mirrors are cooled to about 1~K to prevent the dephasing of atoms due to thermal photons \cite{kuhr2007ultrahigh}.

In the case of the Delayed choice experiment, the atom-cavity interaction and the prob pulse are triggered using a randomly generated binary signal, and the final state of the atom after the interference is co-related with this trigger to analyze the outcome.

\section{Dephasing due to probing}
There are two main ways the non-destructive method of probing the state of the atom affects its phase. When the atom is in state $\ket{1}$ the atom-cavity system experience a global phase shift $\pi$ for every complete cycle of vacuum Rabi oscillation on $\ket{1}\rightarrow\ket{3}$ transition \cite{raimond2001manipulating}. The total phase shift due to N$\times2\pi$ rotations is N$\pi$. 

When the atom in state $\ket{2}$ enters the cavity, it is off-resonant to the cavity and not affected by atom-field coupling. The photons of the probe laser beam present in the cavity shift the $\ket{2}$ atomic level, resulting in a phase shift given by \cite{brune1992manipulation}.

\begin{equation} \label{eqn:phase_off_res}
\phi_{\ket{2}} = \frac{\Omega^2 n}{\delta} t,
\end{equation}

where $\Omega(r)$ is the position-dependent vacuum Rabi coupling, n is the number of photons, t is the interaction time and $\delta$ is the detuning of the photons from the $\ket{2}\rightarrow\ket{3}$ atomic transition ($\delta=\omega_{1\rightarrow 2}$ in this case). Given the identical temporal profile of the probe beam, this term would result in a fixed phase shift for all the iterations.

\section{Conclusion}
Using the method proposed in this article we can probe the ``welcher-Weg" information without destroying the coherence of the atom. Some elements of the experiments are technically challenging, however, similar approaches have been successfully demonstrated in the references cited in this article. This scheme can be extended to perform a ``which state" detection using a single trapped ion in ultra-high finesse optical cavity. 

\section*{Bibliography}


\begin{thebibliography}{18}

\bibitem{bohr1928quantum}
Niels Bohr.
\newblock The quantum postulate and the recent development of atomic theory,
  1928.

\bibitem{feynman1965feynman}
Richard~P Feynman.
\newblock {\em Feynman lectures on physics. Volume 3: Quantum mechancis}.
\newblock 1965.

\bibitem{Wheeler1983quantum}
Law without law.
\newblock In John~Archibald Wheeler and Wojciech~Hubert Zurek, editors, {\em
  Quantum Theory and Measurement}. Princeton University Press, Princeton, NJ,
  1983.

\bibitem{hellmuth1987delayed}
Thomas Hellmuth, Herbert Walther, Arthur Zajonc, and Wolfgang Schleich.
\newblock Delayed-choice experiments in quantum interference.
\newblock {\em Physical Review A}, 35(6):2532, 1987.

\bibitem{durr1998origin}
S~D{\"u}rr, T~Nonn, and G~Rempe.
\newblock Origin of quantum-mechanical complementarity probed by a
  which-way experiment in an atom interferometer.
\newblock {\em Nature}, 395(6697):33--37, 1998.

\bibitem{kim2000delayed}
Yoon-Ho Kim, Rong Yu, Sergei~P Kulik, Yanhua Shih, and Marlan~O Scully.
\newblock Delayed choice quantum eraser.
\newblock {\em Physical Review Letters}, 84(1):1, 2000.

\bibitem{jacques2007experimental}
Vincent Jacques, E~Wu, Fr{\'e}d{\'e}ric Grosshans, Fran{\c{c}}ois Treussart,
  Philippe Grangier, Alain Aspect, and Jean-Fran{\c{c}}ois Roch.
\newblock Experimental realization of wheeler's delayed-choice gedanken
  experiment.
\newblock {\em Science}, 315(5814):966--968, 2007.

\bibitem{manning2015}
Andrew~G Manning, Roman~I Khakimov, Robert~G Dall, and Andrew~G Truscott.
\newblock Wheeler's delayed-choice gedanken experiment with a single atom.
\newblock {\em Nature Physics}, 11(7):539--542, 2015.

\bibitem{storey1993atomic}
Pippa Storey, Matthew Collett, and Daniel Walls.
\newblock Atomic-position resolution by quadrature-field measurement.
\newblock {\em Physical Review A}, 47(1):405, 1993.

\bibitem{gerry1996complementarity}
Christopher~C Gerry.
\newblock Complementarity and quantum erasure with dispersive atom-field
  interactions.
\newblock {\em Physical Review A}, 53(2):1179, 1996.

\bibitem{gleyzes2007quantum}
Sebastien Gleyzes, Stefan Kuhr, Christine Guerlin, Julien Bernu, Samuel
  Deleglise, Ulrich~Busk Hoff, Michel Brune, Jean-Michel Raimond, and Serge
  Haroche.
\newblock Quantum jumps of light recording the birth and death of a photon in a
  cavity.
\newblock {\em Nature}, 446(7133):297--300, 2007.

\bibitem{Hood1998real}
CJ~Hood, MS~Chapman, TW~Lynn, and HJ~Kimble.
\newblock Real-time cavity qed with single atoms.
\newblock {\em Physical review letters}, 80(19):4157, 1998.

\bibitem{assemat2019quantum}
F~Assemat, D~Grosso, A~Signoles, A~Facon, I~Dotsenko, S~Haroche, JM~Raimond,
  M~Brune, and S~Gleyzes.
\newblock Quantum rabi oscillations in coherent and in mesoscopic cat field
  states.
\newblock {\em Physical Review Letters}, 123(14):143605, 2019.

\bibitem{davidovich1993quantum}
L~Davidovich, Abdelhamid Maali, M~Brune, JM~Raimond, and Serge Haroche.
\newblock Quantum switches and nonlocal microwave fields.
\newblock {\em Physical Review Letters}, 71(15):2360, 1993.

\bibitem{sansonetti2006wavelengths}
Jean~E Sansonetti.
\newblock Wavelengths, transition probabilities, and energy levels for the
  spectra of rubidium (rb i through rb xxxvii).
\newblock {\em Journal of physical and chemical reference data},
  35(1):301--421, 2006.

\bibitem{kuhr2007ultrahigh}
Stefan Kuhr, S{\'e}bastien Gleyzes, Christine Guerlin, Julien Bernu, U~Busk
  Hoff, Samuel Del{\'e}glise, Stefano Osnaghi, Michel Brune, J-M Raimond, Serge
  Haroche, et~al.
\newblock Ultrahigh finesse fabry-p{\'e}rot superconducting resonator.
\newblock {\em Applied Physics Letters}, 90(16):164101, 2007.

\bibitem{raimond2001manipulating}
Jean-Michel Raimond, M~Brune, and Serge Haroche.
\newblock Manipulating quantum entanglement with atoms and photons in a cavity.
\newblock {\em Reviews of Modern Physics}, 73(3):565, 2001.

\bibitem{brune1992manipulation}
M~Brune, Serge Haroche, JM~Raimond, Luis Davidovich, and N~Zagury.
\newblock Manipulation of photons in a cavity by dispersive atom-field
  coupling: Quantum-nondemolition measurements and generation of
  schr{\"o}dinger cat states.
\newblock {\em Physical Review A}, 45(7):5193, 1992.

\end{thebibliography}

\end{document}